\documentclass[%
 reprint,
superscriptaddress,
 amsmath,amssymb,
 aps,
prl,
]{revtex4-1}

\usepackage{pdfpages} 
\usepackage{pgffor} 

\makeatletter
\AtBeginDocument{\let\LS@rot\@undefined}
\makeatother

\def\supplementfilename{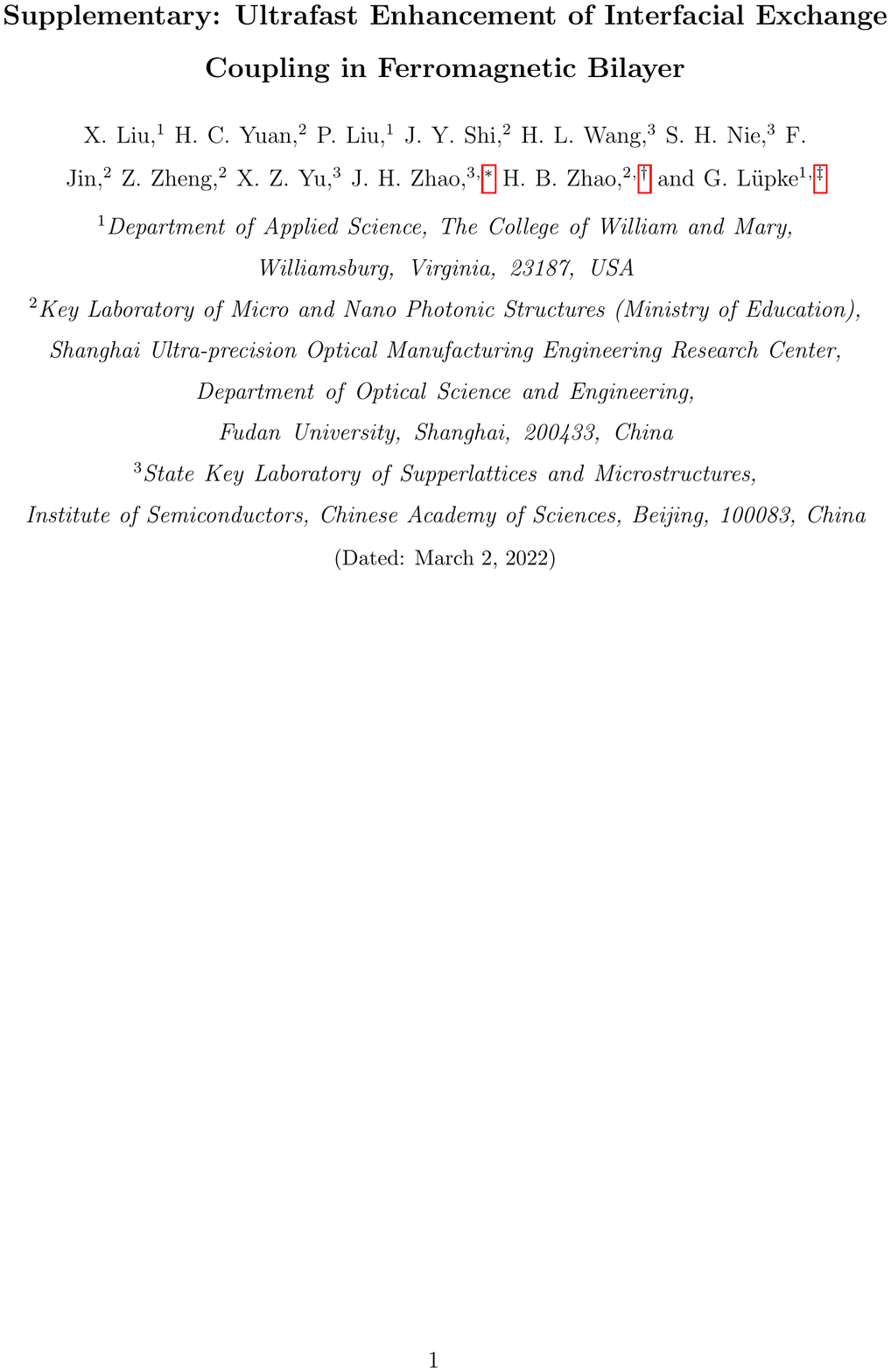}

\pdfximage{\supplementfilename}
\def\numbersupplementpages{\the\pdflastximagepages}

\newif\ifarXiv
\arXivtrue 

\usepackage{graphicx}
\usepackage{epstopdf}
\usepackage{dcolumn}
\usepackage{bm}
\usepackage{gensymb}
\usepackage{upgreek}
\usepackage{hyperref}
\usepackage{graphicx}
\usepackage{epstopdf}
\usepackage{dcolumn}
\usepackage{bm}
\usepackage{gensymb}
\usepackage{upgreek}
\usepackage{hyperref}
\usepackage{lineno}
\usepackage{xr}
\makeatletter
\newcommand*{\addFileDependency}[1]{
  \typeout{(#1)}
  \@addtofilelist{#1}
  \IfFileExists{#1}{}{\typeout{No file #1.}}
}
\makeatother

\begin{document}

\title{Ultrafast Enhancement of Interfacial Exchange Coupling in Ferromagnetic Bilayer}

\author{X. Liu}
\affiliation{Department of Applied Science, The College of William and Mary, Williamsburg, Virginia, 23187, USA}
\author{H. C. Yuan}
\affiliation{Key Laboratory of Micro and Nano Photonic Structures (Ministry of Education), Shanghai Ultra-precision Optical Manufacturing Engineering Research Center, Department of Optical Science and Engineering, Fudan University, Shanghai, 200433, China}
\author{P. Liu}
\affiliation{Department of Applied Science, The College of William and Mary, Williamsburg, Virginia, 23187, USA}
\author{J. Y. Shi}
\affiliation{Key Laboratory of Micro and Nano Photonic Structures (Ministry of Education), Shanghai Ultra-precision Optical Manufacturing Engineering Research Center, Department of Optical Science and Engineering, Fudan University, Shanghai, 200433, China}
\author{H. L. Wang}
\author{S. H. Nie}
\affiliation{State Key Laboratory of Supperlattices and Microstructures, Institute of Semiconductors, Chinese Academy of Sciences, Beijing, 100083, China}
\author{F. Jin}
\author{Z. Zheng}
\affiliation{Key Laboratory of Micro and Nano Photonic Structures (Ministry of Education), Shanghai Ultra-precision Optical Manufacturing Engineering Research Center, Department of Optical Science and Engineering, Fudan University, Shanghai, 200433, China}
\author{X. Z. Yu}
\author{J. H. Zhao}
\thanks{email: jhzhao@red.semi.ac.cn}
\affiliation{State Key Laboratory of Supperlattices and Microstructures, Institute of Semiconductors, Chinese Academy of Sciences, Beijing, 100083, China}
\author{H. B. Zhao}
\thanks{email: hbzhao@fudan.edu.cn}
\affiliation{Key Laboratory of Micro and Nano Photonic Structures (Ministry of Education), Shanghai Ultra-precision Optical Manufacturing Engineering Research Center, Department of Optical Science and Engineering, Fudan University, Shanghai, 200433, China}
\author{G. Lüpke}
\thanks{email: gxluep@wm.edu}
\affiliation{Department of Applied Science, The College of William and Mary, Williamsburg, Virginia, 23187, USA}

\date{\today}

\begin{abstract}
Fast spin manipulation in magnetic heterostructures, where magnetic interactions between different materials often define the functionality of devices, is a key issue in the development of ultrafast spintronics. Although recently developed optical approaches such as ultrafast spin-transfer and spin-orbit torques open new pathways to fast spin manipulation, these processes do not fully utilize the unique possibilities offered by interfacial magnetic coupling effects in ferromagnetic multilayer systems. Here, we experimentally demonstrate ultrafast photo-enhanced interfacial exchange interactions in the ferromagnetic Co$_2$FeAl/(Ga,Mn)As system at low laser fluence levels. The excitation efficiency of Co$_2$FeAl with the (Ga,Mn)As layer is 30-40 times higher than the case with the GaAs layer at 5 K due to a photo-enhanced exchange coupling interaction via photoexcited charge transfer between the two ferromagnetic layers. In addition, the coherent spin precessions persist to room temperature, excluding the drive of photo-enhanced magnetization in the (Ga,Mn)As layer and indicating a proximity-effect-related optical excitation mechanism. The results highlight the importance of considering the range of interfacial exchange interactions in ferromagnetic heterostructures and how these magnetic coupling effects can be utilized for ultrafast, low-power spin manipulation. 
\end{abstract}

\maketitle

\section{\label{sec:Intro}Introduction}

Ultrafast optical control of coherent spin dynamics in ferromagnetic heterostructures is currently of great interest because of its significance in spintronic devices and magnetic recording \cite{1,2,3,4,5,6,7,8,9}. An overwhelming amount of research is devoted to discovering non-thermal, low-power excitation processes to circumvent the disadvantages associate with thermal mechanisms \cite{10,11,12,13,14,15,16}. One promising approach is to make use of the interaction between selectively photo-excited carriers and magnetization since the photo-carrier recombination time is considerably shorter than the thermal dissipation process. Accordingly, it is ideal to utilize the exchange-coupled ferromagnetic systems, as their magnetic properties may respond notably to a small alteration of the exchange-coupling strength \cite{17,18}. Therefore, the combination of a ferromagnetic (FM) metal and a FM semiconductor is specifically favorable for such a purpose of low-power spin manipulation because the FM semiconductor may have pronounced carrier effect on the magnetic properties \cite{19}. Recent studies demonstrated femtosecond optical control of coercivity \cite{20} and complete reversal of magnetic hysteresis loop \cite{21,22,23} in (Ga,Mn)As. Nevertheless, the question is still open whether it is possible to drive FM magnetization at low laser fluence through exchange coupling across metal/semiconductor heterostructure interface.

In this article, ultrafast optical excitation of coherent spin precession is investigated in the exchange-coupled Co$_2$FeAl/(Ga,Mn)As bilayer with time-resolved magneto-optic Kerr effect (TRMOKE). Upon the photoexcitation of electron carriers in (Ga,Mn)As, a nonequilibrium charge current is induced through the heterojunction to the majority spin bands of Co$_2$FeAl, which enhances the interfacial exchange coupling. The photo-excited coherent spin precession persist to room temperature, though the precession amplitude drops significantly around the Curie temperature of (Ga,Mn)As, indicating that the proximity effect plays an essential role in the optical excitation mechanism of coherent spin precession in Co$_2$FeAl/(Ga,Mn)As bilayer. Our results promote the development of low-energy consumption magnetic device concepts for fast spin manipulation.

\section{Results}

\subsection{Outline of the experiments}

\begin{figure}
    \centering
    \advance\leftskip-1cm
    \advance\rightskip-0cm
    \includegraphics[width=0.506\textwidth]{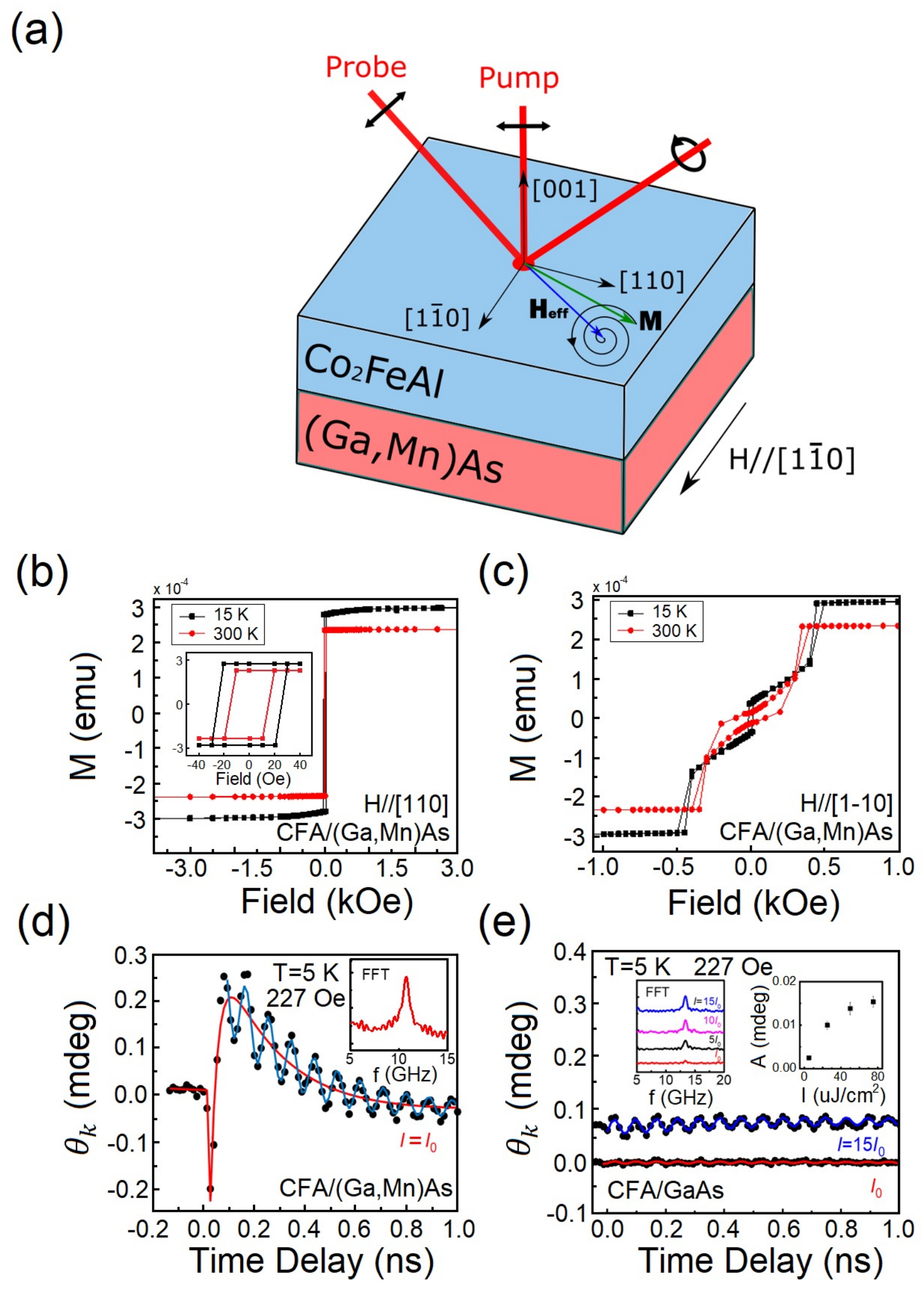}
    \caption{Schematic illustration of TRMOKE experiments, SQUID measurements, and experimental observation of ultrafast enhanced interfacial exchange interaction.(a) Schematic of TRMOKE measurement geometry, depicting the structure of the sample and the magnetization M precessing around the effective field Heff in Co$_2$FeAl/(Ga,Mn)As bilayer in a canted magnetization configuration with H applied along hard axis [1-10]. (b) SQUID measurement in Co$_2$FeAl/(Ga,Mn)As bilayer along the easy axis [110] direction. The inset shows the close-up around zero magnetic field. (c) SQUID measurement in Co$_2$FeAl/(Ga,Mn)As bilayer along the hard axis [1-10] direction. (d) TRMOKE data from Co$_2$FeAl/(Ga,Mn)As bilayer at 5 K, exhibiting the initial demagnetization (-0.41 mdeg) and the subsequent 100-ps magnetization rise (0.46 mdeg) and uniform spin precession. The red and blue curves are fits of LLG equation. The inset shows the FFT analysis of precession frequency. (e) TRMOKE data from Co$_2$FeAl layer without (Ga,Mn)As layer at 5 K. The FFT analysis for precession frequency and the precession amplitude as a function of pump-energy density are shown as insets.
}
    \label{fig:Fig1}
\end{figure}

Our experimental system is a FM exchange-coupled bilayer heterostructure, composed of a near half-metallic Heusler alloy Co$_2$FeAl and a FM semiconductor (Ga,Mn)As (Fig.~\ref{fig:Fig1}), epitaxially grown on GaAs (001) substrate by molecular beam epitaxy (MBE) (see \hyperref[sec:Methods]{Methods}) \cite{24}. The 10-nm thick Co$_2$FeAl layer shows an in-plane uniaxial magnetic anisotropy with an easy axis along the [110] direction (Fig.~\ref{fig:Fig1}(b)], while the 150-nm thick (Ga,Mn)As film reveals an easy axis along the [1-10] direction below the Curie temperature \textit{T$_c$} = 50 K \cite{24}, as revealed by the minor loop in Fig.~\ref{fig:Fig1}(c). 

Due to the strong coupling at the interface and the relatively strong cubic anisotropy constant K1 of Co$_2$FeAl, the interfacial spin alignment of the (Ga,Mn)As layer should be very close to that in the Co$_2$FeAl layer \cite{24}. At low temperature ($T < T_{c}$) a ferromagnetic alignment of Mn spins in the (Ga,Mn)As layer is expected, whereas at high temperatures ($T > T_{c}$) Mn ions extending a few nanometers in the (Ga,Mn)As layer remain spin-polarized due to the ferromagnetic proximity effect \cite{24}. Such a FM metal/semiconductor heterostructure is ideal for examining the interfacial magnetic interactions, due to their different magnetic anisotropy and strong interfacial exchange coupling. Furthermore, one can envision an additional level of optoelectronic control over the underlying exchange interaction due to the existence of spin polarized carriers in the bilayer system. 

\subsection{Observation of ultrafast photo-enhanced exchange interactions}

Coherent spin precessions in Co$_2$FeAl/(Ga,Mn)As bilayer and a reference sample of the Co2FeAl/GaAs bilayer are investigated by near-infrared ($\lambda=800$ nm), low-fluence ($I_{0}=5$ $\mu$J/cm$^{2}$), pump-probe TRMOKE measurements (see Methods) in a canted magnetization configuration where the magnetic field (\textbf{H}) is applied along the hard axis [1-10], as depicted in Fig.~\ref{fig:Fig1}(a). In equilibrium, the magnetization \textbf{M} is along an effective field \textbf{H$_\textup{eff}$}, which is the sum of \textbf{H}, the demagnetizing field, the anisotropy fields, and the exchange-coupling field (see \hyperref[sec:Supplementary]{Supplementary}). The s-polarized incident pump pulses create a short yet strong spin torque $\mathbf{\tau}$(t) in the FM-coupled Co$_2$FeAl/(Ga,Mn)As heterostructure. When $\mathbf{\tau}$ vanishes, the magnetization \textbf{M} is away from its original equilibrium orientation and starts to precess around \textbf{H$_\textup{eff}$}, as depicted in Fig.~\ref{fig:Fig1}(a). The ultrafast spin precession excitation is described by a modified Landau-Lifshitz-Gilbert (LLG) equation with the additional torque term \cite{17}: 
\begin{equation}
    \frac{\partial \mathbf{M}}{\partial t}=-\gamma \left ( \mathbf{M} \times \mathbf{H_{eff}}\right )+\alpha \mathbf{M}\times \frac{\partial \mathbf{M}}{\partial t}+\mathbf{\tau(t)}
    \label{eq:eq1}
\end{equation}
where $\gamma$ is gyromagnetic ratio and $\alpha$ is the Gilbert damping constant.

Figure 1 (d) displays the TRMOKE result from the Co$_2$FeAl/(Ga,Mn)As bilayer at 5 K with the pump pulse fluence $I_{0}=5$ $\mu$J/cm$^{2}$ and magnetic field \textit{H} = 227 Oe. Here, four mutually competing dynamic magnetization processes are observed: (i) an initial $\sim$20 picosecond demagnetization (-0.2 mdeg), (ii) followed by a distinct magnetization rise on a 100 ps time scale (0.2 mdeg), (iii) the magnetization starts precessing in a damped circling way as described by Eq.~\ref{eq:eq1} and (iv) the demagnetization predominates again after the decaying of magnetization enhancement. 

The fast Fourier transform (FFT) of the TRMOKE signal is shown in the inset of Fig.~\ref{fig:Fig1}(d), indicating a precession frequency of around 11 GHz, which is similar to that observed from the single Co$_2$FeAl layer \cite{25}. Besides, the uniform magnetization precession exists up to the room temperature where ferromagnetic order is absent in the (Ga,Mn)As layer. Moreover, since (Ga,Mn)As has much smaller saturated magnetization and magnetic anisotropy, its magnetization precession frequency is much smaller than the observed frequency range.\cite{26,27} Although the (Ga,Mn)As layer contributes to the overall MOKE signal of the bilayer, this contribution does not affect the amplitude of the oscillating component of the TRMOKE signal originating from magnetization precession in the Co$_2$FeAl layer (see \hyperref[sec:Supplementary]{Supplementary Note 6}). Therefore, we ascribe the precession signals observed in the Co$_2$FeAl/(Ga,Mn)As bilayer to the uniform magnetization precession of Co$_2$FeAl, though the overall TRMOKE signal encompasses the contribution of both layers.

Figure~\ref{fig:Fig1}(e)  shows TRMOKE results of Co$_2$FeAl thin film directly grown on GaAs (001) substrate (see \hyperref[sec:Methods]{Methods}), which are remarkably different from those of Co$_2$FeAl/(Ga,Mn)As bilayer. First, very little picosecond demagnetization and no magnetization rise on a 100-ps time scale are observed. Second, almost no coherent spin precession signal can be observed with the same low excitation fluence ($I_{0}=5$ $\mu$J/cm$^{2}$) as the measurement on Co$_2$FeAl/(Ga,Mn)As bilayer, and small oscillations are noticeable only at higher pump energy density, $I=15I_{0}$. Compared with the reference sample Co$_2$FeAl/GaAs, it is pronounced that the excitation efficiency is improved by 30-40 times in the Co$_2$FeAl/(Ga,Mn)As bilayer. In order to elucidate the enhanced optical excitation mechanism, fluence-, field- and temperature-dependent TRMOKE measurements are carried out.

\subsection{Pump-fluence-, field-, and temperature-dependent TRMOKE studies}

\begin{figure}
    \centering
    \advance\leftskip-0.4cm
    \advance\rightskip-0cm
    \includegraphics[width=0.53\textwidth]{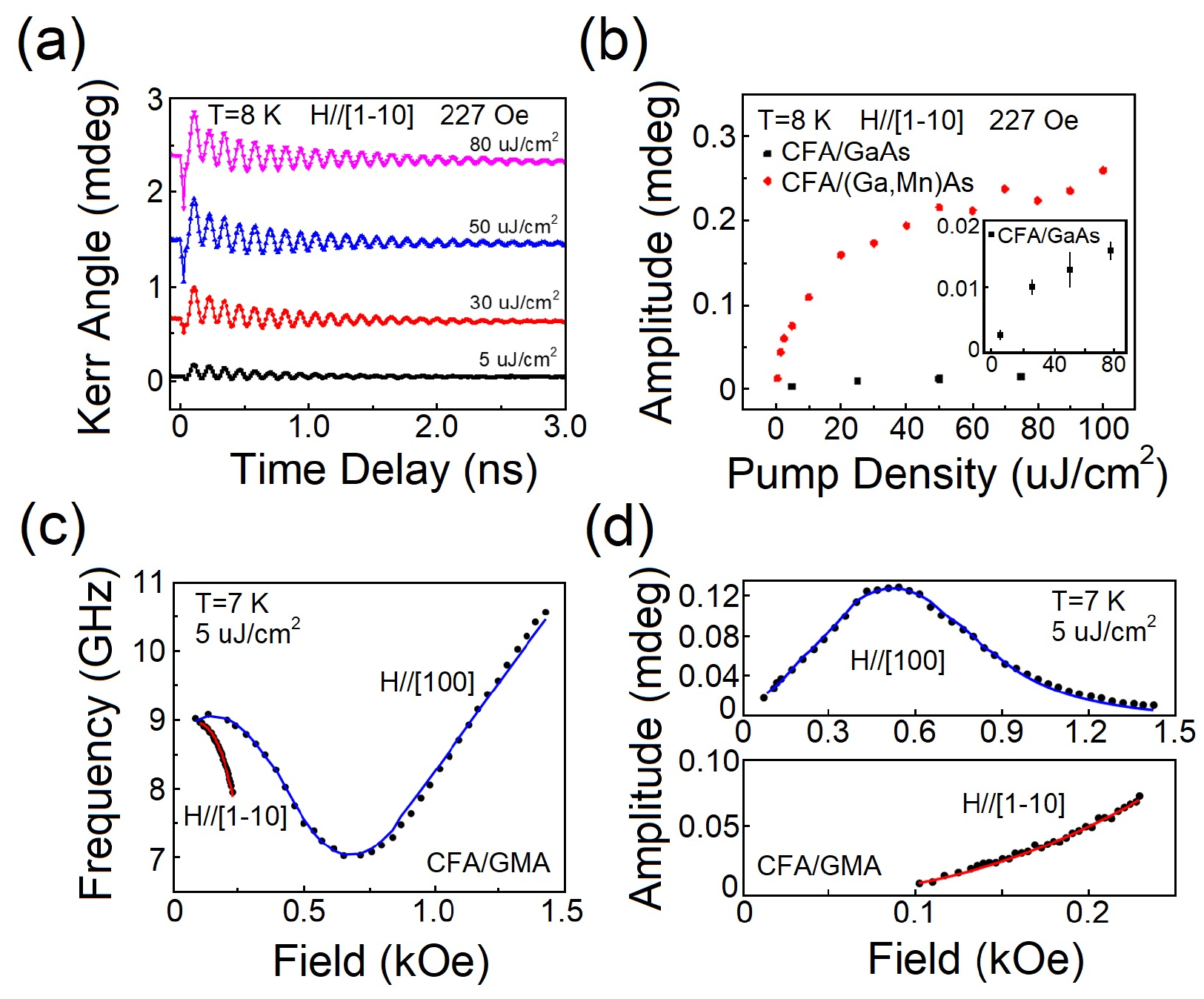}
    \caption{Fluence and field dependent TRMOKE studies. (a) TRMOKE results from Co$_2$FeAl/(Ga,Mn)As bilayer at 8 K with \textbf{H} applied along hard axis [1-10] for different pump-energy densities. (b) Precession amplitude as a function of pump-energy density for Co$_2$FeAl/(Ga,Mn)As (red dots) and Co$_2$FeAl/GaAs (blue dots), respectively. The inset shows close-up of precession amplitude in single Co$_2$FeAl film. (c) Precession frequency as a function of applied field along intermediate axis [100] and hard axis [1-10]. The black dots show the experimental result. The blue and red curves are fits of LLG equation (\hyperref[sec:Supplementary]{Supplementary}). (d) Precession amplitude as a function of the applied magnetic field for [100] (top) and [1-10] direction (bottom), respectively. The black points show the experimental result. The blue and red lines represent fits (\hyperref[sec:Supplementary]{Supplementary}). 
}
    \label{fig:Fig2}
\end{figure}

Figure~\ref{fig:Fig2}(a) and~\ref{fig:Fig2}(b) present TRMOKE results from Co$_2$FeAl/(Ga,Mn)As bilayer as a function of pump-pulse fluence with \textit{H} = 227 Oe applied along the hard axis [1-10]. The precession amplitude \textit{A} increases with increasing pump-pulse fluence and tends to saturate at higher pump fluences, due to thermal effect with high laser power. Figure~\ref{fig:Fig2}(c) and~\ref{fig:Fig2}(d) display the extracted values of precession frequency \textit{f} and amplitude \textit{A}, respectively, as a function of \textit{H} applied along the intermediate axis [100] and hard axis [1-10]. 

The frequency \textit{f} is well fitted with Eq.~\ref{eq:eq1} (red and blue curve) to derive the magnetic anisotropies and the exchange-coupling energy $J_\textup{ex}$ (see \hyperref[sec:Supplementary]{Supplementary Note 2}). At $T$ = 7 K, The magnitude of $J_\textup{ex}$ = 326,056 erg/cm$^{3}$ is comparable to the cubic anisotropy, $K_\textup{1}$ = 290,000 erg/cm$^{3}$, indicating strong ferromagnetic exchange coupling in the Co$_2$FeAl/(Ga,Mn)As bilayer. The exchange-coupling term $J_\textup{ex}$ increases moderately from 300 K to around 50 K, and then increases significantly afterwards (see \hyperref[sec:Supplementary]{Supplementary Note 4}), which suggests a strong exchange-coupling between Co2FeAl and (Ga,Mn)As.   

The simulation of \textit{A} versus \textit{H} (red and blue curve), where \textit{A} is assumed to be proportional to the deviation of equilibrium magnetization $\Delta\varphi$, also agrees well with the experimental data (black dots). At all temperatures, only by considering the enhancement of exchange coupling energies can the experimental data of A versus H satisfy the theoretical model (see \hyperref[sec:Supplementary]{Supplementary Note 5}). This evinces that such strong excitations of magnetization precession originate from the photo-enhancement of interfacial exchange coupling, where the enhancement ratio,, i.e., $\Delta J_\textup{ex}$/$J_\textup{ex}$, is around 20\% at $T$ = 7 K, 10\% at $T$ = 50 K and 5\% at $T$ = 300 K.

\begin{figure}
    \centering
    \advance\leftskip-0.8cm
    \advance\rightskip-0cm
    \includegraphics[width=0.55\textwidth]{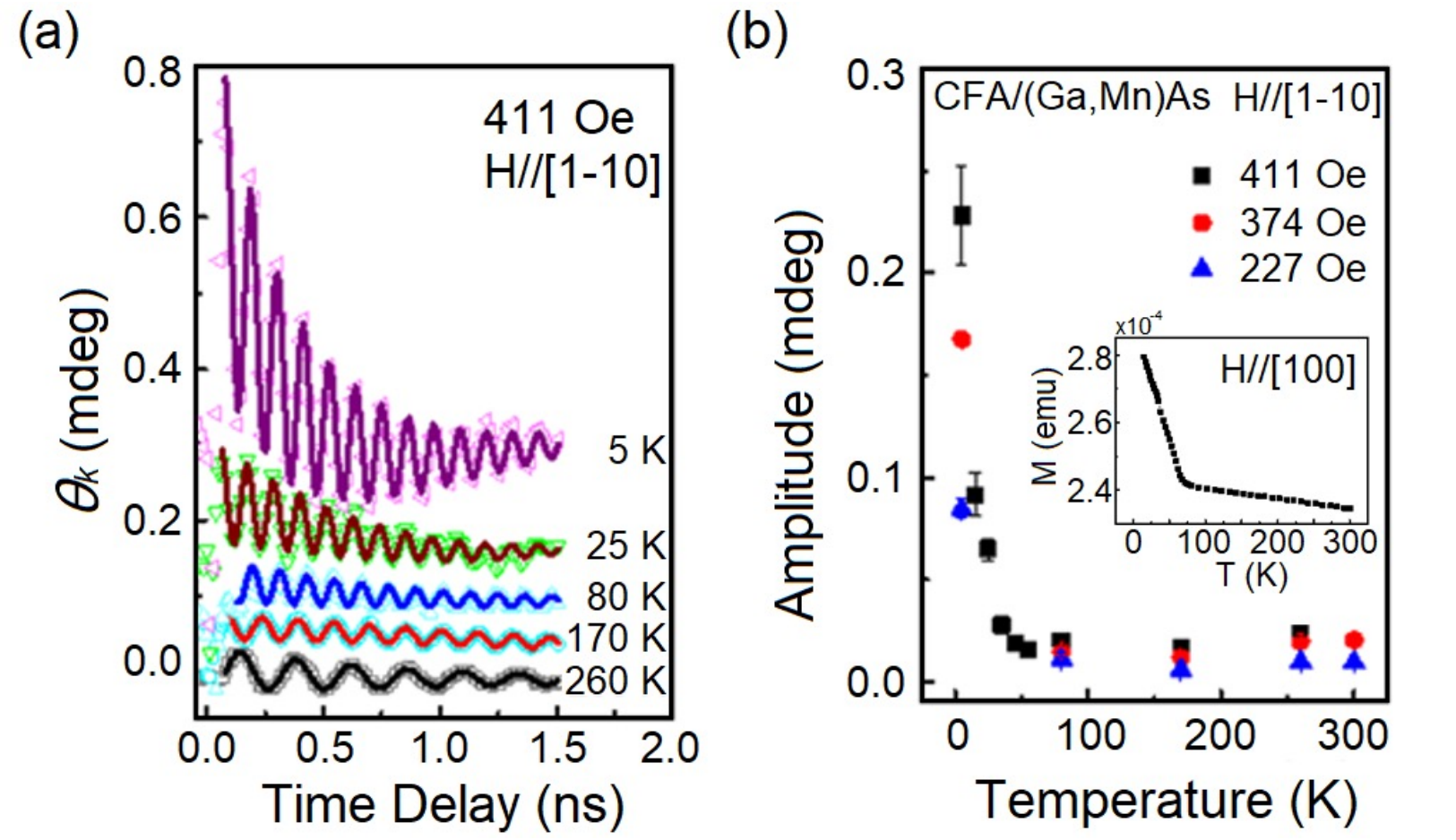}
    \caption{Temperature dependent TRMOKE studies. (a) TRMOKE data from Co$_2$FeAl/(Ga,Mn)As bilayer at different temperatures: 5 K, 25 K, 80 K, 170 K and 260 K. (b) Precession amplitude as a function of temperature at different applied fields: 227 Oe (blue triangles), 374 Oe (red circles) and 411 Oe (black squares). Inset: temperature dependence of static magnetization in Co$_2$FeAl/(Ga,Mn)As bilayer.
}
    \label{fig:Fig3}
\end{figure}

Figure~\ref{fig:Fig3} shows the temperature dependence of the magnetization precession in Co$_2$FeAl/(Ga,Mn)As bilayer with \textit{H} applied along the hard axis [1-10]. The TRMOKE results at \textit{T} = 5 K and 25 K clearly exhibit the photo-induced transient enhancement of magnetization precession amplitude, while at \textit{T} = 80 K, 170 K and 260 K the transient deviation of magnetization is almost absent (Fig.~\ref{fig:Fig3}(a)). Figure~\ref{fig:Fig3}(b) shows the temperature dependence of the precession amplitude \textit{A} for three applied fields: 227 Oe, 374 Oe and 411 Oe. It is prominent that: (i) when the temperature \textit{T} is below the Curie temperature of (Ga,Mn)As $T_{c}$ = 50 K, the precession amplitude \textit{A} rapidly decreases with increasing temperature; and (ii) when \textit{T} is above $T_{c}$, \textit{A} almost remains constant up to room temperature with a small magnitude. In addition, at \textit{T} = 5 K, \textit{A} strongly decreases with decreasing the external field (411 Oe to 227 Oe), while such a reduction is much smaller when \textit{T} $>$ 50 K. On the other hand, the precession amplitude of the reference sample Co$_2$FeAl/GaAs remains unchanged for \textit{T} = 5 K $-$ 300 K, and is 3 $-$ 4 times smaller than that Co$_2$FeAl/(Ga,Mn)As even when \textit{T} $\geqslant$ 50 K. Therefore, these facts suggest that the proximity effect in Co$_2$FeAl/(Ga,Mn)As plays an essential role in the excitation of magnetization precession that is due to the optical modulation of the interfacial exchange coupling.

\section{\label{sec:Discussion}Discussion}

Three aspects are carefully considered to elucidate the pronounced magnetization precession excitation mechanism. First, a comparison among magnetization precessions  of Co$_2$FeAl/(Ga,Mn)As and Co$_2$FeAl/GaAs heterostructures indicates a photo-induced modulation on the interfacial exchange coupling between the Co$_2$FeAl and (Ga,Mn)As layer. If such a modulation is a reduction, we should expect precession amplitudes increase dramatically with temperature increasing near and above 50 K. Because the interfacial exchange coupling will be modulated more significantly at higher temperatures considering it should have a similar trend as frequency (see \hyperref[sec:Supplementary]{Supplementary Note 3}). However, the temperature dependency of precession amplitude is inconsistent with our observations under such circumstances. Second, the precession amplitudes \textit{A} versus applied fields \textit{H} at all temperatures satisfy our theoretical model only when increments are included in interfacial exchange coupling energies $J_\textup{ex}$, rather than magnetic anisotropies or saturation magnetizations. Third, the pump fluence of 5 $\mu$J/cm$^{2}$ is too small to induce heat on the lattice (see \hyperref[sec:Supplementary]{Supplementary Note 7}), making it impossible to modulate the magnetic anisotropies of Co$_2$FeAl with thermal effect. Last, since the temperature dependence of the precession amplitude \textit{A} [Fig.~\ref{fig:Fig3}(b)] follows the same trend with the increase of magnetization, one may think that the photo-enhanced ferromagnetism in the (Ga,Mn)As layer could correlate with the high-efficient excitation of coherent spin precession. However, such a correlation should be insignificant as the photo-enhanced magnetization of (Ga,Mn)As is less than 0.2\% with a similar pump fluence and temperature.\cite{28} Therefore, we attribute the optical excitation mechanism of the observed magnetization precession in the Co$_2$FeAl/(Ga,Mn)As heterostructure to a photo-enhanced interfacial exchange interaction. The mechanism is discussed concretely in the following.

\begin{figure}
    \includegraphics[width=0.4\textwidth]{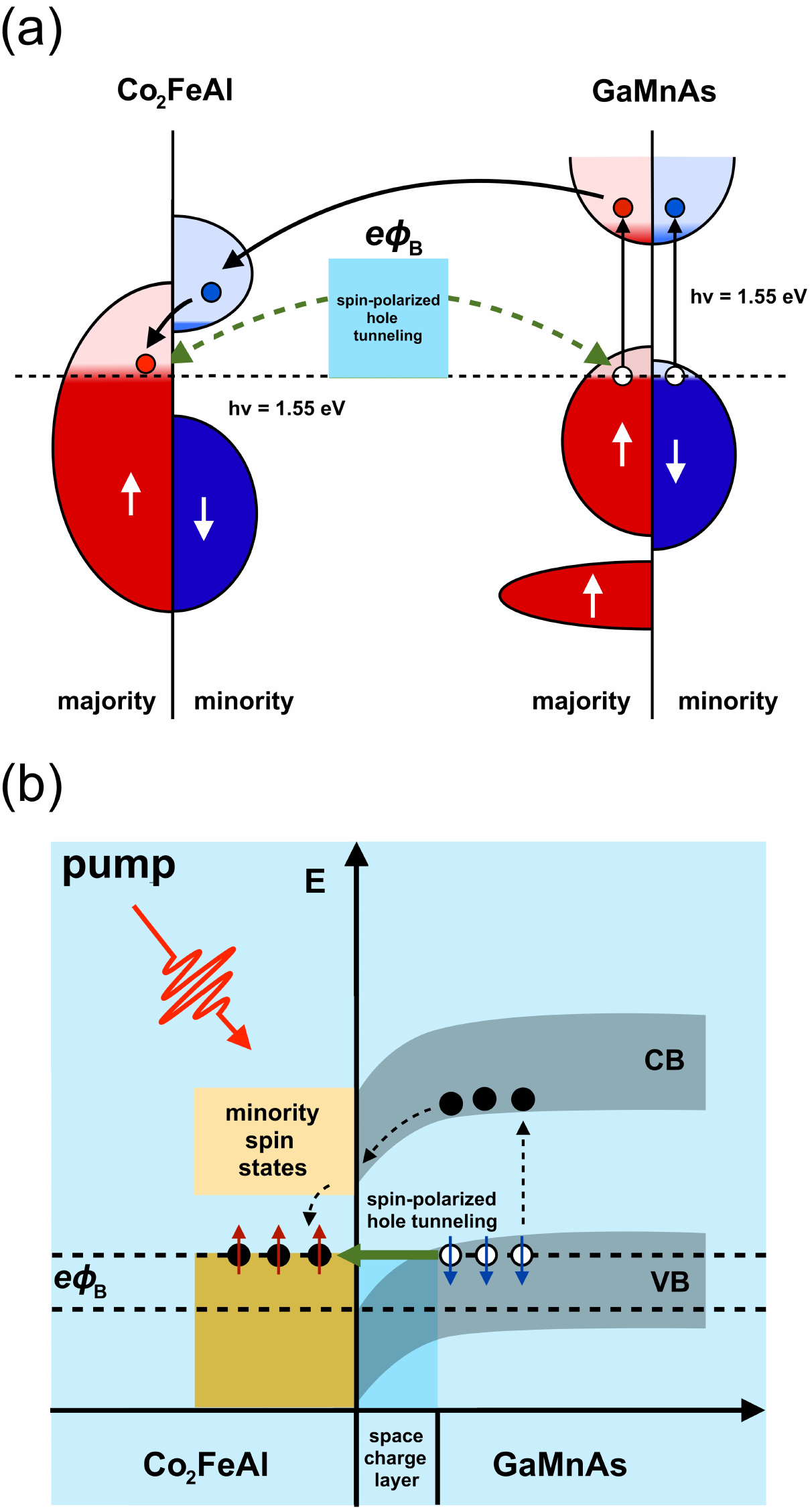}
    \caption{Schematic of the modulation of interfacial exchange coupling via charge transfer to the majority spin states. Transfer and recombination process of photoexcited carriers in the Co$_2$FeAl/(Ga,Mn)As heterostrucutre in (a) the density of state and energy diagram and (b) the corresponding band lineup in real space along the normal to the surface \textit{z}. Red and blue solid circles in (a) corresponds to the majority- and minority-spin electrons, respectively. White and black solid circles in (b) denotes the holes and electrons, respectively.
}
    \label{fig:Fig4}
\end{figure}

A Schottky-barrier space charge layer forms at the heterojunction and the valence band of (Ga,Mn)As is near the Fermi level due to its p-type semiconductor nature, as depicted in  Fig.~\ref{fig:Fig4}(b). Under the low-fluence, near-infrared pump-light excitation, a fast demagnetization process occurs in the Ga,Mn)As layer during $\Delta t<20$ ps due to the rapid spin-flip scattering of the optically generated hot hole carriers with the localized Mn moments, which corresponds to the initial dip in Fig.~\ref{fig:Fig1}(d).  At extended time delays (20 ps $<\Delta t<$ 100 ps), the hole-mediated ferromagnetic ordering enhances as the optically generated, thermalized holes populate the spin-split bands of the (Ga,Mn)As layer. More pronouncedly, the photoexcited excess electrons, as shown in Fig.~\ref{fig:Fig4}b, transfer to the interfacial Co$_2$FeAl layer and fill the majority-spin bands, enhancing the magnetization of Co$_2$FeAl layer at the interface. This results in both a spin current and a charge current that transfers spin angular momentum into the Co$_2$FeAl layer, modulating the FM order of Co$_2$FeAl at the interface and thus enhancing the interfacial exchange coupling. On the other hand, the spin injection to the Co$_2$FeAl layer from the (Ga, Mn)As layer may also produce a spin torque on the Co$_2$FeAl magnetization due to different spin-polarization of the two FM layers. The change of the effective field direction in the Co$_2$FeAl layer gives rise to a transient torque on its magnetization and therefore prominently increases the amplitude of the photo-excited FM spin precession. 

At larger time delays, the magnetization enhancement decays as the excess spin-up electrons in the Co$_2$FeAl layer recombines with the spin-polarized holes in the (Ga,Mn)As through spin-polarized hole tunneling through the Schottky spacer, and the demagnetization in the (Ga,Mn)As layers predominates in the TRMOKE signal as the corresponding recovery process can be as long as 5 ns due to laser-induced heat effect on lattice. \cite{29}

Lastly, although the ferromagnetic ordering in the (Ga,Mn)As layer is almost broken when $T$ $>$ 50 K, the coherent spin precession can still be excited in the Co$_2$FeAl layer with amplitude being 3 $-$ 4 times larger than that of Co$_2$FeAl/GaAs. This consolidates our conclusion that the main contribution to the salient FM precession in Co$_2$FeAl is the modulation of $J_\textup{ex}$ by a photo-excited spin current at the interface rather than the magnetization in the (Ga,Mn)As layer. It is noted that the magnetization precession amplitude \textit{A} exhibits very little variation and remains considerable value with temperatures up to 300 K (Fig.~\ref{fig:Fig3}(b)). Therefore, the spin polarization of the photogenerated holes should correspond to the FM order of (Ga,Mn)As at the interface, which exchange-couple with the Co$_2$FeAl layer via the magnetic proximity effect. In fact, Nie and coworkers have shown that even above the Curie temperature of (Ga,Mn)As, the Mn spins in a 2.11-nm thick interfacial layer remain spin polarized up to 300 K \cite{24}. 

\section{\label{sec:Conclusion}Conclusion}

In summary, the efficiency of coherent spin precession excitation by laser pulses is significantly improved in the Co$_{2}$FeAl/(Ga,Mn)As bilayer, where a large interfacial exchange-coupling establishes. The low-fluence, near-infrared pump-light excitation generates carriers in the (Ga,Mn)As layer, resulting in a charge/spin current to the Co$_{2}$FeAl majority spin states, alternating the FM order of Co$_{2}$FeAl and thus dynamically enhancing the exchange coupling at the interface. The photo-excited coherent spin precession lives up to room temperature, with the amplitude dropping significantly around the Curie temperature of (Ga,Mn)As, indicating that the proximity effect plays an essential role in the optical excitation mechanism of coherent spin precession. Our results highlight the importance of considering the range of interfacial exchange interactions in ferromagnetic heterostructures and promote the development of low-energy consumption magnetic device concepts for fast spin manipulation.

\section{\label{sec:Methods}Methods}

\textbf{Sample fabrication.} Co$_{2}$FeAl/(Ga,Mn)As bilayers are grown on GaAs (001) substrates by molecular-beam epitaxy (MBE) \cite{24}. The thickness of Co$_{2}$FeAl and Ga$_{1-x}$Mn$_{x}$As ($x=0.07$) layers are 10 nm and 150 nm, respectively. The bilayers are capped with 2-nm thick Al layer to avoid oxidation. For the reference, a single 10-nm thick Co$_{2}$FeAl layer is grown on GaAs (001) substrate by MBE. Reflection high-energy electron diffraction (RHEED) patterns, high-resolution double-crystal x-ray diffraction (DCXRD) measurements, and high-resolution cross-sectional transmission electron microscopy (HRTEM) reveal high-quality, single-crystalline, epitaxial growth of the Co$_{2}$FeAl and (Ga,Mn)As films \cite{24}.

\textbf{Magnetic characterization.} Magnetic measurements are conducted from 300 to 5 K in a magnetic field of 1 T using a SQUID magnetometer \cite{24}. The magnetic coupling at the interface between Co2FeAl and (Ga,Mn)As layers is probed with element-resolved XMCD measurements \cite{24}. Ab-initio density functional calculations are performed to support the experimental results \cite{24}.

\textbf{MOKE experiments.} The ferromagnetic magnetization of the exchange-coupled Co$_{2}$FeAl/(Ga,Mn)As bilayer is measured using the longitudinal MOKE setup. The sample is illuminated with p-polarized light and the reflected s-polarized light is detected with a photodiode. The magnetic field is applied along the in-plane [110] or [-110] crystallographic directions. The measurements are conducted from 5 K to above room temperature. 

\textbf{TRMOKE experiments.} For the pump-probe TRMOKE measurements, a Ti:sapphire oscillator laser system is employed, which produces 150-fs pulses at 800-nm wavelength with a repetition rate of 80 MHz. The probe fluence is fixed at $\sim$0.5 $\mu$J/cm$^{2}$ and the pump fluence varies from 0.5 - 100 $\mu$J/cm$^{2}$. The probe pulses ($\lambda=800$ nm) utilize the balanced detection technique with a half-wave plate and Wollaston prism to investigate the transient magnetic state change along longitudinal and polar directions. The measurements are conducted from 5 K to above room temperature.

\textbf{Data analysis and simulations.} The Origin 8.5 software is employed to fit the raw TRMOKE data utilizing the build-in non-linear-fit function. The software provides the uncertainties of precession amplitude \textit{A} and precession frequency \textit{f} from least-square fitting. The Matlab software is utilized to fit the change of \textit{f} and \textit{A} as functions of \textit{H} with least-square methods (see \hyperref[sec:Supplementary]{Supplementary}). The error-propagation method is used to estimate the uncertainties of interfacial exchange-coupling strength and magnetic anisotropy fields.

\section{\label{sec:Contributions}Data Availability}
The data that support the findings of this study are available from the corresponding author upon reasonable request. 

\section{\label{sec:Acknowledge}Acknowledgments}

The work at the College of William and Mary was sponsored by the DOE through Grant No. DEFG02-04ER46127. The work at the Department of Optical Science and Engineering, Fudan University, was supported by the National Natural Science Foundation of China with Grant No. 11774064, National Key Research and Development Program of China (Grant No. 2016YFA0300703), and National Key Basic Research Program (No. 2015CB921403). The work at the State Key Laboratory of Superlattices and Microstructures, Institute of Semiconductors, Chinese Academy of Sciences, was supported by National Natural Science Foundation of China with Grant No. U1632264.

\section{\label{sec:Contributions}Author Contributions}
X. L., P. L., H. C. Y, J. H. Z., H. B. Z., and G. L. designed and analyzed the experiments. H. L. W., S. H. N., J. Y. S., and X. Z. Y. prepared the samples and carried out characterizations using MOKE, RHEED, and SQUID measurements. H. C. Y., J. Y. S., X. L., P. L., and F. J. performed the TRMOKE experiments. X. L., P. L., H. C. Y. and J. Y. S. conducted the data analysis and simulations. All authors discussed the results. X. L., P. L., J. H. Z., H. B. Z., and G. L. wrote the manuscript with contributions from all authors.

\section{\label{sec:Contributions}Competing Interests}
The authors declare no competing financial interests.

\bibliographystyle{naturemag}
\bibliography{ms}

\begin{thebibliography}{10}
\expandafter\ifx\csname url\endcsname\relax
  \def\url#1{\texttt{#1}}\fi
\expandafter\ifx\csname urlprefix\endcsname\relax\def\urlprefix{URL }\fi
\providecommand{\bibinfo}[2]{#2}
\providecommand{\eprint}[2][]{\url{#2}}

\bibitem{1}
\bibinfo{author}{Lambert, C.-H.} \emph{et~al.}
\newblock \bibinfo{title}{{All-optical control of ferromagnetic thin films and
  nanostructures}}.
\newblock \emph{\bibinfo{journal}{Science}} \textbf{\bibinfo{volume}{345}},
  \bibinfo{pages}{1337--1340} (\bibinfo{year}{2014}).

\bibitem{2}
\bibinfo{author}{Guyader, L.~L.} \emph{et~al.}
\newblock \bibinfo{title}{{Nanoscale sub-100 picosecond all-optical
  magnetization switching in GdFeCo microstructures}}.
\newblock \emph{\bibinfo{journal}{Nat. Commun.}} \textbf{\bibinfo{volume}{6}},
  \bibinfo{pages}{5839} (\bibinfo{year}{2015}).

\bibitem{3}
\bibinfo{author}{Rudolf, D.} \emph{et~al.}
\newblock \bibinfo{title}{Ultrafast magnetization enhancement in metallic
  multilayers driven by superdiffusive spin current}.
\newblock \emph{\bibinfo{journal}{Nat. Commun.}} \textbf{\bibinfo{volume}{3}},
  \bibinfo{pages}{1037} (\bibinfo{year}{2012}).

\bibitem{4}
\bibinfo{author}{hoi, G.-M.} \emph{et~al.}
\newblock \bibinfo{title}{Spin current generated by thermally driven ultrafast
  demagnetization}.
\newblock \emph{\bibinfo{journal}{Nat. Commun.}} \textbf{\bibinfo{volume}{5}},
  \bibinfo{pages}{4334} (\bibinfo{year}{2014}).

\bibitem{5}
\bibinfo{author}{Němec, P.} \emph{et~al.}
\newblock \bibinfo{title}{Experimental observation of the optical spin transfer
  torque}.
\newblock \emph{\bibinfo{journal}{Nat. Phys}} \textbf{\bibinfo{volume}{8}},
  \bibinfo{pages}{411–415} (\bibinfo{year}{2012}).

\bibitem{6}
\bibinfo{author}{Tesařová, N.} \emph{et~al.}
\newblock \bibinfo{title}{Experimental observation of the optical spin–orbit
  torque}.
\newblock \emph{\bibinfo{journal}{Nat. Photon.}} \textbf{\bibinfo{volume}{7}},
  \bibinfo{pages}{492–498} (\bibinfo{year}{2013}).

\bibitem{7}
\bibinfo{author}{Chen, J.-Y.} \emph{et~al.}
\newblock \bibinfo{title}{Time-resolved magneto-optical kerr effect of magnetic
  thin films for ultrafast thermal characterization}.
\newblock \emph{\bibinfo{journal}{J. Phys. Chem. Lett.}}
  \textbf{\bibinfo{volume}{7}}, \bibinfo{pages}{2328--2332}
  (\bibinfo{year}{2016}).

\bibitem{8}
\bibinfo{author}{El~Hadri, M.~S.} \emph{et~al.}
\newblock \bibinfo{title}{Electrical characterization of all-optical
  helicity-dependent switching in ferromagnetic hall crosses}.
\newblock \emph{\bibinfo{journal}{Appl. Phys. Lett.}}
  \textbf{\bibinfo{volume}{108}}, \bibinfo{pages}{092405}
  (\bibinfo{year}{2016}).

\bibitem{9}
\bibinfo{author}{Chen, J.-Y.} \emph{et~al.}
\newblock \bibinfo{title}{All-optical switching of magnetic tunnel junctions
  with single subpicosecond laser pulses}.
\newblock \emph{\bibinfo{journal}{Phys. Rev. Applied}}
  \textbf{\bibinfo{volume}{7}}, \bibinfo{pages}{021001} (\bibinfo{year}{2017}).

\bibitem{10}
\bibinfo{author}{Mentink, J.~H.}
\newblock \bibinfo{title}{Manipulating magnetism by ultrafast control of the
  exchange interaction}.
\newblock \emph{\bibinfo{journal}{J. Condens. Matter Phys.}}
  \textbf{\bibinfo{volume}{29}}, \bibinfo{pages}{453001}
  (\bibinfo{year}{2017}).

\bibitem{11}
\bibinfo{author}{Kirilyuk, A.}, \bibinfo{author}{Kimel, A.~V.} \&
  \bibinfo{author}{Rasing, T.}
\newblock \bibinfo{title}{Ultrafast optical manipulation of magnetic order}.
\newblock \emph{\bibinfo{journal}{Rev. Mod. Phys.}}
  \textbf{\bibinfo{volume}{82}}, \bibinfo{pages}{2731--2784}
  (\bibinfo{year}{2010}).

\bibitem{12}
\bibinfo{author}{Hashimoto, Y.}, \bibinfo{author}{Kobayashi, S.} \&
  \bibinfo{author}{Munekata, H.}
\newblock \bibinfo{title}{{Photoinduced precession of magnetization in
  ferromagnetic (Ga,Mn)As}}.
\newblock \emph{\bibinfo{journal}{Phys. Rev. Lett.}}
  \textbf{\bibinfo{volume}{100}}, \bibinfo{pages}{067202}
  (\bibinfo{year}{2008}).

\bibitem{13}
\bibinfo{author}{Radu, I.} \emph{et~al.}
\newblock \bibinfo{title}{Transient ferromagnetic-like state mediating
  ultrafast reversal of antiferromagnetically coupled spins}.
\newblock \emph{\bibinfo{journal}{Nature}} \textbf{\bibinfo{volume}{472}},
  \bibinfo{pages}{205–208} (\bibinfo{year}{2011}).

\bibitem{14}
\bibinfo{author}{Mathias, S.} \emph{et~al.}
\newblock \bibinfo{title}{Probing the timescale of the exchange interaction in
  a ferromagnetic alloy}.
\newblock \emph{\bibinfo{journal}{Proc. Natl. Acad. Sci.}}
  \textbf{\bibinfo{volume}{109}}, \bibinfo{pages}{4792--4797}
  (\bibinfo{year}{2012}).

\bibitem{15}
\bibinfo{author}{Stupakiewicz, A.} \emph{et~al.}
\newblock \bibinfo{title}{Ultrafast nonthermal photo-magnetic recording in a
  transparent medium}.
\newblock \emph{\bibinfo{journal}{Nature}} \textbf{\bibinfo{volume}{542}},
  \bibinfo{pages}{71–74} (\bibinfo{year}{2017}).

\bibitem{16}
\bibinfo{author}{Mikhaylovskiy, R.} \emph{et~al.}
\newblock \bibinfo{title}{Ultrafast optical modification of exchange
  interactions in iron oxides}.
\newblock \emph{\bibinfo{journal}{Nat. Commun.}} \textbf{\bibinfo{volume}{6}},
  \bibinfo{pages}{8190} (\bibinfo{year}{2015}).

\bibitem{17}
\bibinfo{author}{Ma, X.} \emph{et~al.}
\newblock \bibinfo{title}{Ultrafast spin exchange-coupling torque via
  photo-excited charge-transfer processes}.
\newblock \emph{\bibinfo{journal}{Nat. Commun.}} \textbf{\bibinfo{volume}{6}},
  \bibinfo{pages}{8800} (\bibinfo{year}{2015}).

\bibitem{18}
\bibinfo{author}{Fan, Y.} \emph{et~al.}
\newblock \bibinfo{title}{Photoinduced spin angular momentum transfer into an
  antiferromagnetic insulator}.
\newblock \emph{\bibinfo{journal}{Phys. Rev. B}} \textbf{\bibinfo{volume}{89}},
  \bibinfo{pages}{094428} (\bibinfo{year}{2014}).

\bibitem{19}
\bibinfo{author}{Matsubara, M.} \emph{et~al.}
\newblock \bibinfo{title}{Ultrafast optical tuning of ferromagnetism via the
  carrier density}.
\newblock \emph{\bibinfo{journal}{Nat. Commun.}} \textbf{\bibinfo{volume}{6}},
  \bibinfo{pages}{6724} (\bibinfo{year}{2015}).

\bibitem{20}
\bibinfo{author}{Hall, K.~C.} \emph{et~al.}
\newblock \bibinfo{title}{{Ultrafast optical control of coercivity in GaMnAs}}.
\newblock \emph{\bibinfo{journal}{Appl. Phys. Lett.}}
  \textbf{\bibinfo{volume}{93}}, \bibinfo{pages}{032504}
  (\bibinfo{year}{2008}).

\bibitem{21}
\bibinfo{author}{Astakhov, G.~V.} \emph{et~al.}
\newblock \bibinfo{title}{{Magnetization manipulation in (Ga,Mn)As by
  subpicosecond optical excitation}}.
\newblock \emph{\bibinfo{journal}{Appl. Phys. Lett.}}
  \textbf{\bibinfo{volume}{86}}, \bibinfo{pages}{152506}
  (\bibinfo{year}{2005}).

\bibitem{22}
\bibinfo{author}{Wang, J.} \emph{et~al.}
\newblock \bibinfo{title}{{Memory effects in photoinduced femtosecond
  magnetization rotation in ferromagnetic GaMnAs}}.
\newblock \emph{\bibinfo{journal}{Appl. Phys. Lett.}}
  \textbf{\bibinfo{volume}{94}}, \bibinfo{pages}{021101}
  (\bibinfo{year}{2009}).

\bibitem{23}
\bibinfo{author}{Zhu, Y.} \emph{et~al.}
\newblock \bibinfo{title}{{Ultrafast dynamics of four-state magnetization
  reversal in (Ga,Mn)As}}.
\newblock \emph{\bibinfo{journal}{Appl. Phys. Lett.}}
  \textbf{\bibinfo{volume}{95}}, \bibinfo{pages}{052108}
  (\bibinfo{year}{2009}).

\bibitem{24}
\bibinfo{author}{Nie, S.~H.} \emph{et~al.}
\newblock \bibinfo{title}{{Ferromagnetic interfacial interaction and the
  proximity effect in a
  ${\mathrm{Co}}_{2}\mathrm{FeAl}/(\mathrm{Ga},\mathrm{Mn})\mathrm{As}$
  bilayer}}.
\newblock \emph{\bibinfo{journal}{Phys. Rev. Lett.}}
  \textbf{\bibinfo{volume}{111}}, \bibinfo{pages}{027203}
  (\bibinfo{year}{2013}).

\bibitem{25}
\bibinfo{author}{Yuan, H.~C.} \emph{et~al.}
\newblock \bibinfo{title}{{Different temperature scaling of strain-induced
  magneto-crystalline anisotropy and Gilbert damping in Co$_{2}$FeAl film
  epitaxied on GaAs}}.
\newblock \emph{\bibinfo{journal}{Appl. Phys. Lett.}}
  \textbf{\bibinfo{volume}{105}}, \bibinfo{pages}{072413}
  (\bibinfo{year}{2014}).

\bibitem{26}
\bibinfo{author}{Rozkotová, E.} \emph{et~al.}
\newblock \bibinfo{title}{{Light-induced magnetization precession in GaMnAs}}.
\newblock \emph{\bibinfo{journal}{Applied Physics Letters}}
  \textbf{\bibinfo{volume}{92}}, \bibinfo{pages}{122507}
  (\bibinfo{year}{2008}).

\bibitem{27}
\bibinfo{author}{Qi, J.} \emph{et~al.}
\newblock \bibinfo{title}{{Coherent magnetization precession in GaMnAs induced
  by ultrafast optical excitation}}.
\newblock \emph{\bibinfo{journal}{Appl. Phys. Lett.}}
  \textbf{\bibinfo{volume}{91}}, \bibinfo{pages}{112506}
  (\bibinfo{year}{2007}).

\bibitem{28}
\bibinfo{author}{Wang, J.} \emph{et~al.}
\newblock \bibinfo{title}{{Ultrafast Enhancement of Ferromagnetism via
  Photoexcited Holes in GaMnAs}}.
\newblock \emph{\bibinfo{journal}{Phys. Rev. Lett.}}
  \textbf{\bibinfo{volume}{98}}, \bibinfo{pages}{217401}
  (\bibinfo{year}{2007}).

\bibitem{29}
\bibinfo{author}{Zahn, J.~P.} \emph{et~al.}
\newblock \bibinfo{title}{{Ultrafast studies of carrier and magnetization
  dynamics in GaMnAs}}.
\newblock \emph{\bibinfo{journal}{J. Appl. Phys.}}
  \textbf{\bibinfo{volume}{107}}, \bibinfo{pages}{033908}
  (\bibinfo{year}{2010}).

\end{thebibliography}

\ifarXiv
    

\section*{Supplementary material}

**SUPPLEMENTAY EQUATIONS**

**Equation S1: Magnetic free energy ($E$)**

$$E = -M_s H \cos(\varphi_M - \varphi_H) - (2\pi M_s^2 + K_\perp) \sin^2 \varphi_\perp + K_u \sin^2 \varphi_M$$
$$+ \frac{K_1}{4} \sin^2 2\varphi_M + K_{ud} \sin(\varphi_M - \varphi_{ud}) + K_{ra} \cos^2(\varphi_M - \varphi_H) + J_1 M_s M_s' \cos\theta, \quad (S1)$$

The last term $J_{\text{ex}} = J_1 M_s M_s' \cos\theta$ represents the contribution from the exchange coupling effect, where $M_s$ is the saturation magnetization in the Co$_2$FeAl layer and $M_s'$ is the saturation magnetization in the (Ga,Mn)As layer. $J_1$ is the exchange-coupling stiffness and $\theta$ is the angle between the magnetization directions in Co$_2$FeAl and (Ga,Mn)As. $H$ is the applied field. $\varphi_M$, $\varphi_H$, $\varphi_\perp$ and $\varphi_{ud}$ are the angles of magnetization, applied field, perpendicular and unidirectional anisotropy, respectively. $K_\perp$, $K_u$, $K_1$, $K_{ud}$ and $K_{ra}$ are the out-of-plane, in-plane uniaxial, crystalline cubic, unidirectional and rotatable magnetic anisotropies, respectively.

**Equation S2: Spin precession frequency ($f$)**

$$f = \frac{\gamma}{2\pi} (H_a \cdot H_b)^{\frac{1}{2}}, \quad (S2)$$

where

$$H_a = H \cos(\varphi_M - \varphi_H) + H_u \cos 2\varphi_M + H_1 \cos 4\varphi_M$$
$$- \frac{H_{ud}}{2} \sin(\varphi_M - \varphi_{ud}) + H_{ra} \cos 2(\varphi_M - \varphi_H) - J_1 M_s' \cos\theta,$$

and

$$H_b = H \cos(\varphi_M - \varphi_H) + 4\pi M_{eff} - H_u \sin^2 \varphi_M + \frac{H_1}{4}(3 + \cos 4\varphi_M)$$
$$- \frac{H_{ud}}{2} \sin(\varphi_M - \varphi_{ud}) + H_{ra} \cos^2(\varphi_M - \varphi_H) - J_1 M_s' \cos\theta,$$

with $M_{eff} = M_s + \frac{K_\perp}{2\pi M_s}$, $H_u = \frac{2K_u}{M_s}$, $H_1 = \frac{2K_1}{M_s}$, $H_{ud} = \frac{2K_{ud}}{M_s}$, $H_{ra} = \frac{2K_{ra}}{M_s}$ and gyromagnetic ratio $\gamma = 1.76 \times 10^7$ Hz/Oe.

**Equation S3: Simulation of precession amplitude (A)**

$$A = a\Delta\varphi, \quad (S3)$$

where $a$ is the scaling pre-factor for the conversion from $\Delta\varphi$ to $A$, and $\Delta\varphi$ is the change in magnetization direction.

**SUPPLEMENTAY NOTES**

**Note 1: Static MOKE Measurement**

Figure S1 presents the temperature dependence of the static MOKE measurement performed along [110] direction in $Co_2FeAl/(Ga,Mn)As$ bilayer (# L1288) and $Co_2FeAl/GaAs$ structure (# L688). The laser diode is set at 670 nm. We notice in the $Co_2FeAl/(Ga,Mn)As$ bilayer that the Kerr angle first rises and then drops with decreasing temperature (Fig. S1(a)),

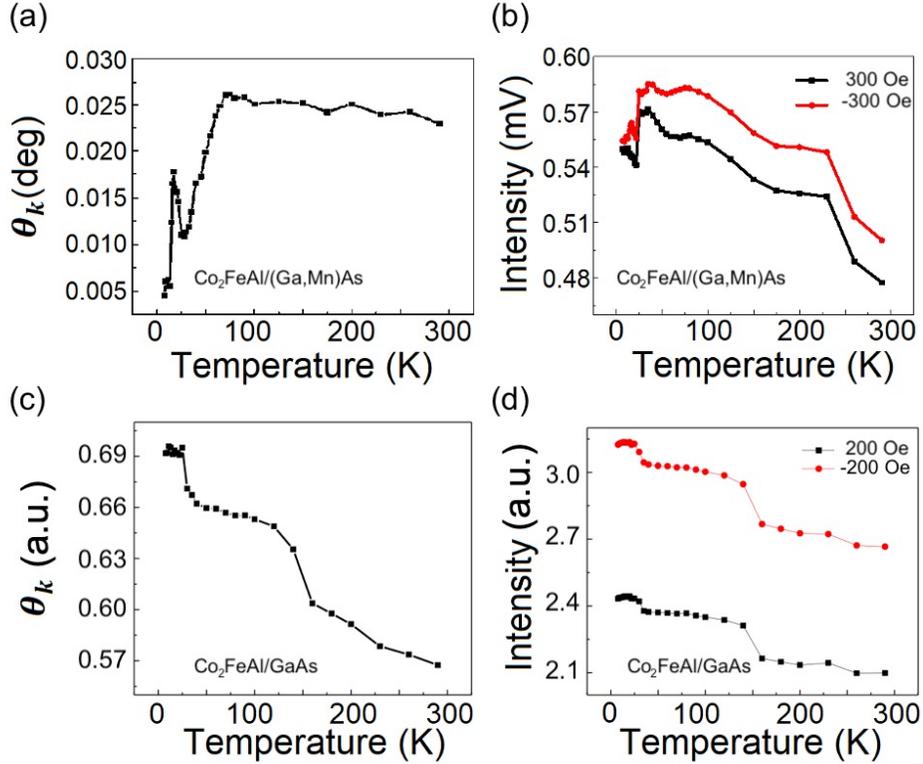

FIG. S1. Temperature dependence of static MOKE measurement from $Co_2FeAl/(Ga,Mn)As$ bilayer: (a) Kerr angle and (b) intensity and $Co_2FeAl/GaAs$ bilayer: (c) Kerr angle and (d) intensity.

while the intensity decreases with decreasing temperature below the Curie temperature $T_c$ = 50 K of (Ga,Mn)As (Fig. S1(b)). In contrast, the MOKE signal shows a slight increase in the Co$_2$eAl/GaAs structure when the temperature decreases from 300 K to 5 K (Fig. S1(c)), and similar behavior is observed for the intensity versus temperature as well (Fig. S1(d)).

**Note 2: Extraction of Anisotropy Fields and Exchange Coupling Strength**

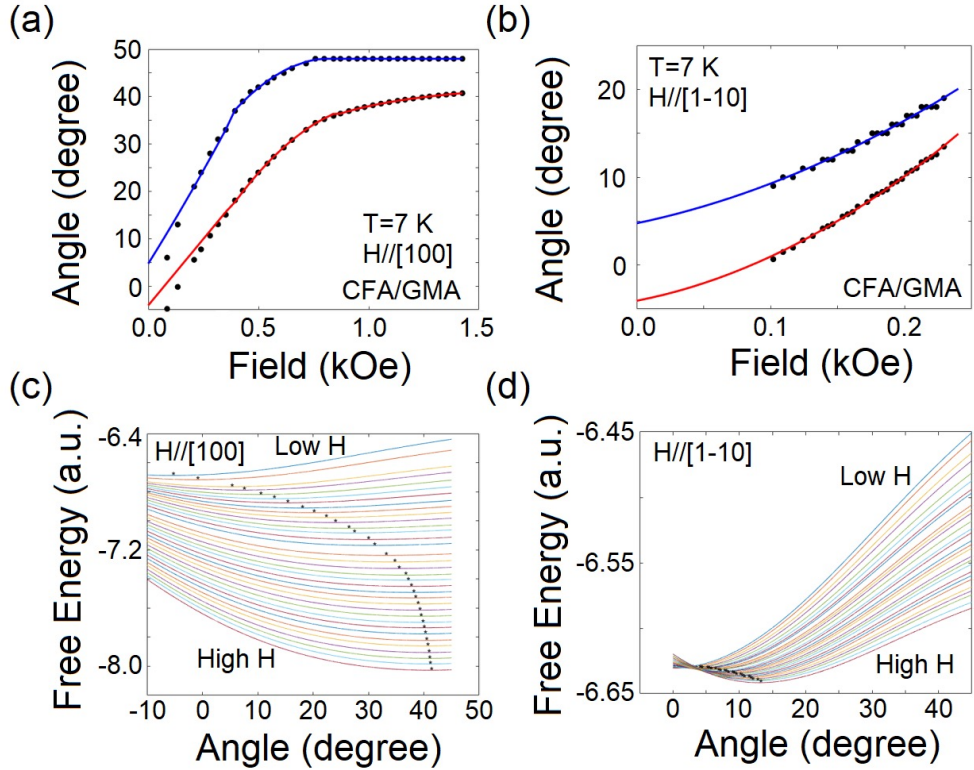

FIG. S2. Angle of magnetization direction in Co$_2$FeAl and (Ga,Mn)As as a function of external magnetic field applied along (a) [100] direction and (b) [1-10] direction. The angle of magnetization is given with respect to crystallographic axis [110]. The red curve represents the approximated polynomial functions for the Co$_2$FeAl layer while the blue curve represents the approximated polynomial functions for the (Ga,Mn)As layer. Magnetic free energy as a function of magnetization direction in Co$_2$FeAl layer with the external magnetic field applied along (c) [100] direction and (d) [1-10] direction. Curves in different colors correspond to different external fields. The black dot in each curve presents the minimum of magnetic free energy for each applied field.

The magnetic anisotropy fields are obtained by analyzing the field dependence of the mag-

netization precession frequency at temperature $T = 7$ K. First, by minimizing the magnetic free energy $E$ (Eq. S1), the angle of magnetization directions is determined as a function of external field for a certain range of anisotropy fields. Then we can obtain the value of magnetic anisotropy fields by fitting the precession frequency $f$ with Equation S2.

Figure S2(a) and S2(b) present the angle of magnetization direction for [100] and [1-10] directions, respectively. The angles of magnetization direction of $Co_2FeAl$ layer (upper dots) and (Ga,Mn)As layer (lower dots) are calculated by minimizing the magnetic free energy (Figs. S2(c) and S2(d)) and are quoted with respect to [110] direction. They are used further to fit the precession amplitudes (Fig. 2(d), main text), which is discussed below in Note 4. For the magnetic field applied along [100] direction, the angle of magnetization direction in (Ga,Mn)As layer reaches saturation at 48 degrees for applied fields above 800 Oe, which is close to the direction of the applied field. This indicates that the direction of magnetization in (Ga,Mn)As structure is mainly driven by the external field. The blue and red curves are approximated from the angle of magnetizations in both layers by polynomial functions. Then these smooth curves are used to fit the field dependence of precession frequency, as shown in Fig. 2(c) of main text. Although there is a little deviation between the dots and the fitted curves at the low field part for [100] direction, the error can be neglected because the magnetic free energy has a broad minimum at low field (Fig. S2(c)). The best fitting results are $K_u = 0$ erg/cm$^3$, $K_1 = 290,000$ erg/cm$^3$, $K_{ud} = 155,000$ erg/cm3 along [100] direction, $K_{ra} = 30,000$ erg/cm$^3$, $M_{eff} = 880$ emu/cm$^3$, $M_s = 1,127$ emu/cm$^3$ and $J_{\text{ex}} = 326,056$ erg/cm$^3$.

**Note 3: Temperature dependence of precession frequency**

Figure S3 shows the temperature dependency of the magnetization precession frequency under different applied fields: 227 Oe, 374 Oe and 411 Oe. The precession frequencies increase significantly from 300 K to 100 K and barely change below 50 K, revealing a similar static exchange coupling energy trend.

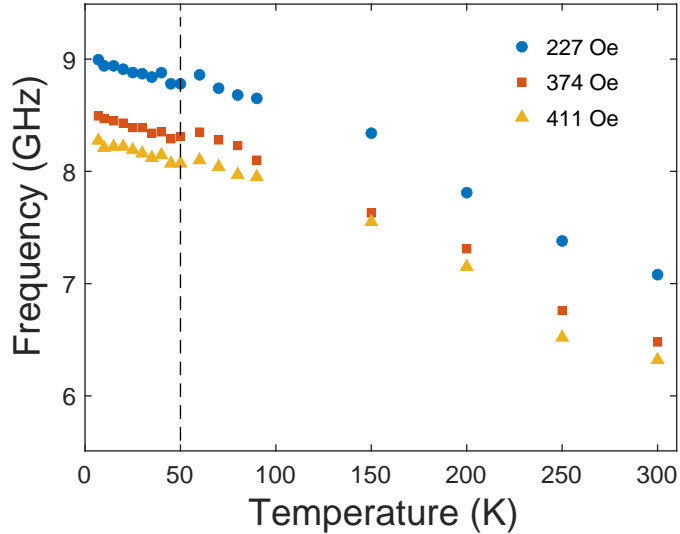

FIG. S3. Temperature dependence of the magnetization precession frequency under different applied fields——228 Oe, 412 Oe and 596 Oe, respectively.

**Note 4: Temperature dependence of magnetic parameters**

In order to look into the origin of the efficient excitation mechanism below the Curie temperature of the (Ga,Mn)As layer, we simulated the spin precession frequency and amplitude by Equation S2 and S3 through the same procedure mentioned in Note 2 at different temperature. The obtained temperature dependence of the exchange-coupling effect, unidirectional field, rotatable anisotropy field and cubic anisotropy field from 7 K to 300 K are shown in Table S1. Since there's no obvious temperature dependence of the uniaxial anisotropy field $K_u$, effective magnetic field $M_{eff}$ and saturation magnetization $M_s$ in the $Co_2FeAl$ layer, they are considered as constant value ($K_u$ = 4.4×103 erg/cm$^3$, $M_{eff}$ = 915 emu/cm$^3$ and $M_s$ = 1,130 emu/cm$^3$). Note that there might be no uniform magnetization in the (Ga,Mn)As layer when it is exchange-coupled to the $Co_2FeAl$ layer. Because the ferromagnetic magnetization in the $Co_2FeAl$ layer applies a torque on the Mn spins at the interface, which leads to reorientation of the magnetic moments with increasing depth in the (Ga,Mn)As layer [1]. Hence, we consider the exchange coupling term as a whole in the simulation. The model yields satisfactory agreement with our experimental results, which are consistent with the observations of the transient enhancement of total magnetization induced in (Ga,Mn)As by low pump excitation on a 100-ps time scale, as reported by Wang's

| $T$ (K) | $J_{\text{ex}}$ (erg/cm$^3$) | $K_{\text{ud}}$ (erg/cm$^3$) | $K_{\text{ra}}$ (erg/cm$^3$) | $K_1$ (erg/cm$^3$) |
|---|---|---|---|---|
| 7 | 326,056 | 155,000 | 30,000 | 290,000 |
| 10 | 279,212 | 150,000 | 30,000 | 290,000 |
| 15 | 264,963 | 150,000 | 30,000 | 290,000 |
| 20 | 249,886 | 160,000 | 15,000 | 290,000 |
| 25 | 255,190 | 120,000 | 10,000 | 300,000 |
| 30 | 245,125 | 130,000 | 10,000 | 290,000 |
| 35 | 255,190 | 120,000 | 10,000 | 290,000 |
| 40 | 235,824 | 100,000 | 5000 | 290,000 |
| 45 | 221,333 | 100,000 | 5000 | 290,000 |
| 50 | 241,276 | 100,000 | 5000 | 290,000 |
| 60 | 238,560 | 80,000 | 0 | 300,000 |
| 70 | 238,560 | 80,000 | 0 | 300,000 |
| 80 | 247,800 | 90,000 | 0 | 290,000 |
| 90 | 228,850 | 80,000 | 0 | 280,000 |
| 150 | 209,937 | 65,000 | 0 | 250,000 |
| 200 | 159,920 | 80,000 | 0 | 220,000 |
| 250 | 179,946 | 60,000 | 0 | 220,000 |
| 300 | 144,957 | 65,000 | 0 | 190,000 |

TABLE S1. Temperature dependence of exchange-coupling effect, unidirectional field, rotatable anisotropy field and cubic anisotropy field obtained from TRMOKE measurement of the Co$_2$FeAl/(Ga,Mn)As bilayers.

group [2]. Here the saturation magnetization in the (Ga,Mn)As layer ($M_s = 38.4$ emu/cm$^3$ at 5 K) is obtained by static measurement (Fig. 1(c), main text). As a consequence, one can see the increasing temperature weakens the interfacial exchange-coupling strength $J_{\text{ex}}$ and the unidirectional anisotropy field $K_{ud}$ while the other anisotropy fields are relatively stable below the Curie temperature of (Ga,Mn)As. Furthermore, the uniaxial anisotropy field has little contribution to the magnetic response, so does the rotatable anisotropy field which disappears completely as the temperature goes above $T_c = 50$ K. This helps us exclude the influence from the modulation in the magnetic anisotropy fields and focus on the contribution from the change in the interfacial exchange coupling strength.

**Note 5: Simulation of precession amplitude**

We investigate the relationship between the field-dependent precession amplitude $A$ and precession frequency f from 7 K to 300 K. As mentioned before, the angle of magnetization

7

| $T$ (K) | $\Delta J_{\text{ex}}/J_{\text{ex}}$ | $\Delta K_{\text{ud}}/K_{\text{ud}}$ | $\Delta K_{\text{ra}}/K_{\text{ra}}$ | $\Delta K_1/K_1$ |
|---|---|---|---|---|
| 7 | 20% | 0 | 0 | 0 |
| 10 | 19.5% | 0 | 0 | 0 |
| 15 | 18% | 0 | 0 | 0 |
| 20 | 16% | 0 | 0 | 0 |
| 25 | 16.5% | 0 | 0 | 0 |
| 30 | 14% | 0 | 0 | 0 |
| 35 | 13% | 0 | 0 | 0 |
| 40 | 11% | 0 | 0 | 0 |
| 45 | 11% | 0 | 0 | 0 |
| 50 | 12.5% | 0 | 0 | 0 |
| 60 | 11% | 0 | 0 | 0 |
| 70 | 10% | 0 | 0 | 0 |
| 80 | 9.5% | 0 | 0 | 0 |
| 90 | 8% | 0 | 0 | 0 |
| 150 | 7% | 0 | 0 | 0 |
| 200 | 4% | 0 | 0 | 0 |
| 250 | 5.5% | 0 | 0 | 0 |
| 300 | 5% | 0 | 0 | 0 |

TABLE S2. Temperature dependence of exchange-coupling effect, unidirectional field, rotatable anisotropy field and cubic anisotropy field obtained from TRMOKE measurement of the Co$_2$FeAl/(Ga,Mn)As bilayers.

direction is derived by minimizing the magnetic free energy using Equation S1. We then approximated the precession amplitude $A$ by the change $\Delta\varphi$ in magnetization direction due to the transient change of interfacial exchange interaction, unidirectional field, rotatable anisotropy field, and cubic anisotropy field as expressed in Equation S3 (Table S2). For the data curves at all temperatures, $A$ is fitted well by $\Delta\varphi$, as shown in Fig. 2(d) of main text.

**Note 6: Contribution from the (Ga,Mn)As layer**

Although the (Ga,Mn)As layer contributes to the magnetic response of the Co$_2$FeAl/(Ga,Mn)As bilayer, this contribution shouldn't affect the modulated MOKE signal induced by the magnetization precession in the Co$_2$FeAl layer, as can be proved by the following derivation. For balanced detection and above Curie temperature (considering p-polarized incident light),

$$I_0 = \left[\frac{\sqrt{2}}{2}\left(E_p^{CFA} + E_p^{GMA}\right) + \frac{\sqrt{2}}{2}E_s^{CFA}\right]^2 - \left[\frac{\sqrt{2}}{2}\left(E_p^{CFA} + E_p^{GMA}\right) - \frac{\sqrt{2}}{2}E_s^{CFA}\right]^2$$
$$= 2\left(E_p^{CFA} + E_p^{GMA}\right) \times E_s^{CFA},$$

$$I = \left[\frac{\sqrt{2}}{2}\left(E_p^{CFA} + E_p^{GMA}\right) + \frac{\sqrt{2}}{2}\left(E_s^{CFA} + \Delta E_s^{CFA}\right)\right]^2 - \left[\frac{\sqrt{2}}{2}\left(E_p^{CFA} + E_p^{GMA}\right) - \frac{\sqrt{2}}{2}\left(E_s^{CFA} + \Delta E_s^{CFA}\right)\right]^2$$
$$= 2\left(E_p^{CFA} + E_p^{GMA}\right) \times \left(E_s^{CFA} + \Delta E_s^{CFA}\right),$$

$$I - I_0 = 2\left(E_p^{CFA} + E_p^{GMA}\right) \times \Delta E_s^{CFA},$$

and below Curie temperature,

$$I'_0 = \left[\frac{\sqrt{2}}{2}\left(E_p^{CFA} + E_p^{GMA}\right) + \frac{\sqrt{2}}{2}\left(E_s^{CFA} + E_s^{GMA}\right)\right]^2 - \left[\frac{\sqrt{2}}{2}\left(E_p^{CFA} + E_p^{GMA}\right) - \frac{\sqrt{2}}{2}\left(E_s^{CFA} + E_s^{GMA}\right)\right]^2$$
$$= 2\left(E_p^{CFA} + E_p^{GMA}\right) \times \left(E_s^{CFA} + E_s^{GMA}\right),$$

$$I'_0 = \left[\frac{\sqrt{2}}{2}\left(E_p^{CFA} + E_p^{GMA}\right) + \frac{\sqrt{2}}{2}\left(E_s^{CFA} + \Delta E_s^{CFA} + E_s^{GMA}\right)\right]^2 - \left[\frac{\sqrt{2}}{2}\left(E_p^{CFA} + E_p^{GMA}\right) - \frac{\sqrt{2}}{2}\left(E_s^{CFA} + \Delta E_s^{CFA} + E_s^{GMA}\right)\right]^2$$
$$= 2\left(E_p^{CFA} + E_p^{GMA}\right) \times \left(E_s^{CFA} + \Delta E_s^{CFA} + E_s^{GMA}\right),$$

$$I' - I'_0 = 2\left(E_p^{CFA} + E_p^{GMA}\right) \times \Delta E_s^{CFA}.$$

One can see that $E_s^{GMA}$ has no contribution to the modulated MOKE signal at temperature below the Curie temperature of (Ga,Mn)As. Note that in the above calculation, we assume the same phase of reflected electric fields from both layers. Although the actual phases are different, it will not change the conclusion.

### Note 7: Photo-induced thermal effect estimation

The 2 nm thin $Al_2O_3$ layer with a high transmission at 800 nm (greater than 95%) is considered to be transparent. The estimated reflectivity of $Co_2FeAl$ is R = 50%.[3] The estimated refraction and extinction coefficient for $Co_2FeAl$ thin film is $n = 6$ and $\kappa = 3$,[4] respectively, and the absorption coefficient at 800 nm is derived as $\alpha_1 = 4\pi\kappa/\lambda \approx 0.047$ $nm^{-1}$. The refraction and extinction coefficient for (Ga,Mn)As thin film is $n = 3.5$ and $\kappa = 0.08$,[5, 6] respectively, and the absorption coefficient at 800 nm is derived as $\alpha_2 = 4\pi\kappa/\lambda \approx$

9

0.001 nm$^{-1}$. Within the first several hundred femtoseconds or first few picoseconds after the laser pulse, electrons are excited to a high temperature and then the energy is transferred to the lattice through electron-electron and electron-phonon scattering. The estimation of the temperature increase is given by

$$(1-R)IAe^{-\alpha_1 d_1}\alpha_2 \int_0^{d_2} e^{-\alpha_2 x}\, dx = C_v \frac{\rho A d_2}{M}\Delta T.$$

where $\rho = 5.3$ g/cm$^3$ is the density of GaAs, $C_v = 6.3$ mJg$^{-1}$K$^{-1}$ is the special heat capacity of GaAs at 5 K [7], and the thickness of Co$_2$FeAl and (Ga,Mn)As layers is $d_1 = 10$ nm and $d_1 = 150$ nm. The estimated increment of temperature with a pump fluence of 5 $\mu$J/cm$^2$ is $\Delta T = 0.5$ K (0.8 K if considering full reflection). Considering the heat dissipation, the laser-induced heat effect is completely negligible.

---

* email: jhzhao@red.semi.ac.cn

† email: hbzhao@fudan.edu.cn

‡ email: gxluep@wm.edu

[1] Stiles, M. D. & McMichael, R. D. Model for exchange bias in polycrystalline ferromagnet-antiferromagnet bilayers. *Phys. Rev. B* **59**, 3722–3733 (1999).

[2] Wang, J. *et al.* Ultrafast Enhancement of Ferromagnetism via Photoexcited Holes in GaMnAs. *Phys. Rev. Lett.* **98**, 217401 (2007).

[3] Jain, V. K. *et al.* Electronic structure, magnetic and optical properties of quaternary Fe$_{2-x}$Co$_x$MnAl Heusler alloys. *J. Mater. Sci.* **52**, 6800–6811 (2017).

[4] Jain, R. *et al.* Electronic structure, magnetic and optical properties of Co$_2$Ti$_Z$ (Z = B, Al, Ga, In) Heusler alloys. *J. Magn. Magn. Mater.* **448**, 278–286 (2018).

[5] Teradaa, H., Ohyaband, S. & Tanakac, M. Intrinsic magneto-optical spectra of GaMnAs. *Appl. Phys. Lett.* **106**, 222406 (2015).

[6] Aspnes, D. E. *et al.* Optical properties of Al$_x$Ga$_{1-x}$As. *J. Appl. Phys.* **60**, 754 (1998).

[7] Cetas, T. C., Tilford, C. R. & Swenson, C. A. Specific Heats of Cu, GaAs, GaSb, InAs, and InSb. *Phys. Rev.* **174**, 835 (1968).


\fi

\end{document}